\documentclass{article}
\usepackage{cite}

\begin{document}

\date{}
\title{Algebraic treatment of the Pais-Uhlenbeck oscillator and its PT-variant}
\author{Francisco M. Fern\'{a}ndez \thanks{%
E-mail: fernande@quimica.unlp.edu.ar} \\
INIFTA (CONICET, UNLP), Divisi\'on Qu\'imica Te\'orica\\
Blvd. 113 S/N, Sucursal 4, Casilla de Correo 16, 1900 La Plata, Argentina}
\maketitle

\begin{abstract}
The algebraic method enables one to study the properties of the spectrum of
a quadratic Hamiltonian through the mathematical properties of a matrix
representation called regular or adjoint. This matrix exhibits exceptional
points where it becomes defective and can be written in canonical Jordan
form. It is shown that any quadratic function of $K$ coordinates and $K$
momenta leads to a $2K$ differential equation for those dynamical variables.
We illustrate all these features of the algebraic method by means of the
Pais-Uhlenbeck oscillator and its PT-variant. We also consider a trivial
quantization of the fourth-order differential equation for the
quantum-mechanical dynamical variables.
\end{abstract}

\section{Introduction}

\label{sec:intro}

Since the seminal paper on generalizations of the field equations to
equations of higher order by Pais and Uhlenbeck\cite{PU50} there has been
several attempts at a suitable quantization of fourth-order dynamical
differential equations\cite{MD05,S06,M07,BM08a,BM08b,S09,M10,M11}. Most of
the papers are based on the Pais-Uhlenbeck oscillator\cite
{PU50,S06,M07,BM08a,BM08b,S09,M11} but there are also other model candidates%
\cite{M10,LM12}. All these oscillators are quadratic functions of the
coordinates and their conjugate momenta. In principle, any such Hamiltonian
leads to a fourth-order dynamical differential equation\cite{F16b}. In the
discussion of the just mentioned quantization, concepts like exceptional
points\cite{S09} (see\cite{HS90,H00,HH01,H04} for a detailed discussion),
breaking of commutation relations in the equal-frequency case\cite{MD05},
apparent reduction of the number of degrees of freedom\cite{MD05,S06} and
Jordan matrices\cite{MD05,S06,M07,BM08b,S09} appear over an over again.

The algebraic method is extremely useful for the analysis of the
mathematical properties of quadratic Hamiltonians\cite{F15a,F16,F16b}. It
consists of associating each quadratic function of $K$ coordinates and their
$K$ conjugate momenta with a $2K\times 2K$ matrix. The main features of the
spectrum of the Hamiltonian emerge from the mathematical properties of its
adjoint or regular matrix representation\cite{G74,FC96}.

The purpose of this paper is to apply the algebraic method to the
Pais-Uhlenbeck oscillator\cite{MD05,S06,M07,BM08a,BM08b,S09,M10,M11} and its
complex-PT variant\cite{BM08a,BM08b,S09} and show that exceptional points,
breaking of commutation relations, apparent reduction of degrees of freedom
and Jordan matrices appear in a quite natural and straightforward way. At
the same time we show the advantage of using the algebraic method for the
analysis of this kind of oscillators.

In section~\ref{sec:algebraic_method} we review the main equations of the
algebraic method for Hamiltonians that are quadratic functions of $K$
coordinates and their conjugate momenta and show that any such operator
leads to a differential equation of order $2K$ for each dynamical variable.
In section~\ref{sec:Pais_Uhl} we apply the algebraic method to the
Pais-Uhlenbeck oscillator and its PT-symmetric variant. In section~\ref
{sec:trivial} we consider a trivial quantization of the fourth-order
differential equation. Finally, in section~\ref{sec:conclusions} we
summarize the main results of the paper and draw conclusions.

\section{The algebraic method}

\label{sec:algebraic_method}

The algebraic method is particularly useful for the analysis of the spectrum
of quantum-mechanical Hamiltonians that are quadratic functions of the
coordinates and their conjugate momenta:
\begin{equation}
H=\sum_{i=1}^{2K}\sum_{j=1}^{2K}\gamma _{ij}O_{i}O_{j},
\label{eq:H_quadratic}
\end{equation}
where $S_{2K}=\left\{ O_{1},O_{2},\ldots ,O_{2K}\right\} =\left\{
x_{1},x_{2},\ldots ,x_{K},p_{1},p_{2},\ldots ,p_{K}\right\} $, $%
[x_{m},p_{n}]=i\delta _{mn}$, and $[x_{m},x_{n}]=[p_{m},p_{n}]=0$. The
method may also be applied to classical problems\cite{F02} but we do not
discuss this case here. Since the Hamiltonian (\ref{eq:H_quadratic})
satisfies the closure commutation relations
\begin{equation}
\lbrack H,O_{i}]=\sum_{j=1}^{2K}H_{ji}O_{j},\;i=1,2,\ldots ,2K,
\label{eq:[H,Oi]}
\end{equation}
we can define a $2K\times 2K$ matrix $\mathbf{H}$ with elements $H_{ij}$
that is commonly known as the adjoint or regular matrix representation of $H$
in the operator basis set $S_{2K}$\cite{G74,FC96}. In order to make this
paper sufficiently self-contained, in this section we review the main
results of our earlier papers\cite{F16,F16b}.

Because of equation (\ref{eq:[H,Oi]}) it is possible to find an operator of
the form
\begin{equation}
Z=\sum_{i=1}^{N}c_{i}O_{i},  \label{eq:Z}
\end{equation}
such that
\begin{equation}
\lbrack H,Z]=\lambda Z,  \label{eq:[H,Z]}
\end{equation}
where $\lambda $ is a complex number. If $\left| \psi \right\rangle $ is an
eigenvector of $H$ with eigenvalue $E$ then
\begin{equation}
HZ\left| \psi \right\rangle =ZH\left| \psi \right\rangle +\lambda Z\left|
\psi \right\rangle =(E+\lambda )Z\left| \psi \right\rangle ;
\label{eq:HZ|Psi>}
\end{equation}
that is to say, $Z\left| \psi \right\rangle $ is an eigenvector of $H$ with
eigenvalue $E+\lambda $. It follows from equations (\ref{eq:[H,Oi]}), (\ref
{eq:Z}) and (\ref{eq:[H,Z]}) that the coefficients $c_{i}$ are solutions to
\begin{equation}
(\mathbf{H}-\lambda \mathbf{I})\mathbf{C}=0,  \label{eq:(H-lambda_I)C=0}
\end{equation}
where $\mathbf{I}$ is the $2K\times 2K$ identity matrix and $\mathbf{C}$ is
a $2K\times 1$ column matrix with elements $c_{i}$. Equation (\ref
{eq:(H-lambda_I)C=0}) admits nontrivial solutions for those values of $%
\lambda $ that are roots of
\begin{equation}
P(\lambda )=\det (\mathbf{H}-\lambda \mathbf{I})=0.  \label{eq:charpoly}
\end{equation}
Clearly, the eigenvalues $\lambda _{i}$, $i=1,2,\ldots ,2K$ are the natural
frequencies of $H$ (differences between pairs of eigenvalues of this
operator). This matrix representation is not normal
\begin{equation}
\mathbf{HH}^{\dagger }\neq \mathbf{H}^{\dagger }\mathbf{H,}
\label{eq:H_nonnormal}
\end{equation}
and for this reason it may be defective or non-diagonalizable (it may not
have a complete basis set of eigenvectors).

We can also define the matrix $\mathbf{U}$ with elements $U_{ij}$ given by
\begin{equation}
\lbrack O_{i},O_{j}]=U_{ij}\hat{1},  \label{eq:[Oi,Oj]}
\end{equation}
where $\hat{1}$ is the identity operator that we will omit from now on. This
matrix can be written as
\begin{equation}
\mathbf{U}=i\left(
\begin{array}{ll}
\mathbf{0} & \mathbf{I} \\
-\mathbf{I} & \mathbf{0}
\end{array}
\right) ,  \label{eq:U_matrix}
\end{equation}
where $\mathbf{0}$ and $\mathbf{I}$ are the $K\times K$ zero and identity
matrices, respectively. Note that $\mathbf{U}^{\dagger }=\mathbf{U}^{-1}=%
\mathbf{U}$. The matrices $\mathbf{H}$ and $\mathbf{U}$ are related by
\begin{equation}
\mathbf{H}=(\mathbf{\gamma }+\mathbf{\gamma }^{t})\mathbf{U,}
\label{eq:H=gamma.U}
\end{equation}
where $\mathbf{\gamma }$ is the $2K\times 2K$ matrix with elements $\gamma
_{ij}$.

The well known Jacobi identity $%
[O_{k},[H,O_{i}]]+[O_{i},[O_{k},H]]+[H,[O_{i},O_{k}]]=0$ leads to $%
[O_{k},[H,O_{i}]]=[O_{i},[H,O_{k}]]$. Therefore, it follows from the latter
equation, (\ref{eq:[H,Oi]}) and (\ref{eq:[Oi,Oj]}) that $(\mathbf{UH})^{t}=%
\mathbf{H}^{t}\mathbf{U}^{t}=-\mathbf{H}^{t}\mathbf{U}=\mathbf{UH},$which
leads to $\mathbf{UH}^{t}\mathbf{U}=-\mathbf{H}$. We can thus prove that $%
\det (\mathbf{H}+\lambda \mathbf{I})=P(-\lambda )=0$; that is to say: if $%
\lambda $ is an eigenvalue then $-\lambda $ is also an eigenvalue.\ In other
words, $P(\lambda )$ is a polynomial function of $\lambda ^{2}$. We also
have
\begin{eqnarray}
\mathbf{H}^{t}\mathbf{UC} &=&-\lambda \mathbf{UC}  \nonumber \\
\mathbf{H}^{\dagger }\mathbf{UC}^{*} &=&-\lambda ^{*}\mathbf{UC}^{*}.
\label{eq:H^t_UC}
\end{eqnarray}

If $\mathbf{\gamma }^{\dagger }=\mathbf{\gamma }$ the quadratic Hamiltonian (%
\ref{eq:H_quadratic}) is Hermitian and $\mathbf{H}$ is $\mathbf{U}$-pseudo
Hermitian\cite{F16}:
\begin{equation}
\mathbf{H}^{\dagger }=\mathbf{UHU},  \label{eq:pseudo_hermit}
\end{equation}
(for a more detailed discussion of quasi-Hermiticity or pseudo-Hermiticity
see\cite{P43,SGH92,M02a,M02b,M02c}). In this case $\mathbf{\gamma +\gamma }%
^{t}=\mathbf{\gamma +\gamma }^{*}=2\Re \mathbf{\gamma }$. Besides, it
follows from $[H,Z]^{\dagger }=[Z^{\dagger },H]=\lambda ^{*}Z^{\dagger }$
that if $Z$ is solution to equation (\ref{eq:[H,Z]}), then $Z^{\dagger }$ is
solution to
\begin{equation}
\lbrack H,Z^{\dagger }]=-\lambda ^{*}Z^{\dagger }.  \label{eq:[H,Z^dagger]}
\end{equation}
In other words: both $\lambda $ and $-\lambda ^{*}$ are roots of $P(\lambda
)=0$.

For every eigenvalue $\lambda _{i}$ we construct the operator
\begin{equation}
Z_{i}=\sum_{j=1}^{2K}c_{ij}O_{j}.
\end{equation}
For convenience we label the eigenvalues in such a way that $\lambda
_{j}=-\lambda _{2K-j+1}$, $j=1,2,\ldots ,K$, and when they are real we
organize them in the following way:
\begin{equation}
\lambda _{1}<\lambda _{2}<\ldots <\lambda _{K}<0<\lambda _{K+1}<\ldots
<\lambda _{2K}.
\end{equation}

If we take into account that $[H,Z_{i}Z_{j}]=\left( \lambda _{i}+\lambda
_{j}\right) Z_{i}Z_{j}$ then we conclude that
\begin{equation}
\lbrack H,[Z_{i},Z_{j}]]=\left( \lambda _{i}+\lambda _{j}\right)
[Z_{i},Z_{j}]=0,
\end{equation}
which tells us that $Z_{i}$ and $Z_{j}$ commute when $\lambda _{i}+\lambda
_{j}\neq 0$. If $[Z_{j},Z_{2K-j+1}]=\sigma _{j}\neq 0$ for all $j=1,2,\ldots
,K$ then we can write $H$ in the following way
\begin{equation}
H=-\sum_{j=1}^{K}\frac{\lambda _{j}}{\sigma _{j}}Z_{2K-j+1}Z_{j}+E_{0}.
\label{eq:H(Z_j)}
\end{equation}
If $\psi _{0}$ is a vector in the Hilbert space where $H$ is defined that
satisfies
\begin{equation}
Z_{j}\psi _{0}=0,\;j=1,2,\ldots ,K,
\end{equation}
then $H\psi _{0}=E_{0}\psi _{0}.$

Consider the time-evolution of the dynamical variables
\begin{equation}
O_{j}(t)=e^{itH}O_{j}e^{-itH},
\end{equation}
and their equations of motion
\begin{equation}
\dot{O}_{j}(t)=ie^{itH}[H,O_{j}]e^{-itH}=i\sum_{k=1}^{2K}H_{kj}O_{k}(t).
\end{equation}
If we define the row vector $\mathbf{O}(t)=\left( O_{1}(t)\,O_{2}(t)\,\ldots
\,O_{2K}(t)\right) $ then we have the matrix differential equation $\mathbf{%
\dot{O}}(t)=i\mathbf{O}(t)\mathbf{H}$ with solution
\begin{equation}
\mathbf{O}(t)=\mathbf{O}e^{it\mathbf{H}},\;\mathbf{O}=\mathbf{O}(0).
\end{equation}
Since $P(\mathbf{H})=0$ then
\begin{equation}
P\left( -i\frac{d}{dt}\right) \mathbf{O}(t)=\mathbf{O}P(\mathbf{H})e^{it%
\mathbf{H}}=0,
\end{equation}
gives us a differential equation of order $2K$ for the dynamical variables.
Obviously, $Z_{j}(t)=e^{it\lambda _{j}}Z_{j}$, $j=1,2,\ldots ,2K$, satisfies
this equation. It clearly tells us that any Hamiltonian that is a quadratic
function of $K$ coordinates ant their conjugate momenta leads to a
differential equation of order $2K$ for any such dynamical variable. In
particular, for $K=2$ we have $P(\lambda )=\left( \lambda ^{2}-\lambda
_{1}^{2}\right) \left( \lambda ^{2}-\lambda _{2}^{2}\right) =0$ and the
fourth-order differential equation
\begin{equation}
\frac{d^{4}}{dt^{4}}q+\left( \lambda _{1}^{2}+\lambda _{2}^{2}\right) \frac{%
d^{2}}{dt^{2}}q+\lambda _{1}^{2}\lambda _{2}^{2}q=0,  \label{eq:diff_eq_f-o}
\end{equation}
where $q\in S_{4}$. In principle, any pair of coupled oscillators may be a
candidate for the quantization of a fourth-order differential equation like (%
\ref{eq:diff_eq_f-o}). Some of them have already been discussed and analysed%
\cite{S06,M07,BM08a,BM08b,S09,M10,M11,LM12}, and many more can be proposed.

\section{The Pais-Uhlenbeck oscillator}

\label{sec:Pais_Uhl}

In this section we consider the \textit{standard} Pais-Uhlenbeck oscillator%
\cite{PU50,S06,M07,BM08a,BM08b,S09,M11}
\begin{equation}
H=\frac{1}{2}p_{x}^{2}+xp_{y}+\frac{\omega _{1}^{2}+\omega _{2}^{2}}{2}x^{2}-%
\frac{\omega _{1}^{2}\omega _{2}^{2}}{2}y^{2},  \label{eq:H_PU}
\end{equation}
as well as its PT-symmetric modification\cite{BM08a,BM08b}
\begin{equation}
H=\frac{1}{2}p_{x}^{2}-ixp_{y}+\frac{\omega _{1}^{2}+\omega _{2}^{2}}{2}%
x^{2}+\frac{\omega _{1}^{2}\omega _{2}^{2}}{2}y^{2}.  \label{eq:H_PU_mod}
\end{equation}
The latter exhibits two antiunitary symmetries\cite{W60} given by $%
A_{1}:(x,y,p_{x},p_{y})\rightarrow (-x,-y,p_{x},p_{y})$, $%
A_{2}:(x,y,p_{x},p_{y})\rightarrow (x,y,-p_{x},-p_{y})$ that satisfy $%
A_{q}=A_{q}^{-1}$, $q=1,2$, and $A_{q}iA_{q}^{-1}=-i$. Since $%
A_{q}HA_{q}^{-1}=H$ we have $HA_{q}\left| \psi \right\rangle =A_{q}H\left|
\psi \right\rangle =A_{q}E\left| \psi \right\rangle =E^{*}\left| \psi
\right\rangle $. If the antiunitary symmetry is exact $A_{q}\left| \psi
\right\rangle =a_{q}\left| \psi \right\rangle $ (with $a_{q}$ being a
complex number) then the eigenvalue $E$ is real. The antiunitary symmetry $%
A_{1}$ was chosen by Bender and Manheim\cite{BM08b} in their analysis of the
quantization of the fourth-order differential equation.

In what follows we consider the somewhat more general oscillator
\begin{equation}
H=\frac{1}{2}p_{x}^{2}+axp_{y}+\frac{\omega _{1}^{2}+\omega _{2}^{2}}{2}%
x^{2}+b\frac{\omega _{1}^{2}\omega _{2}^{2}}{2}y^{2},  \label{eq:H_PU_gen}
\end{equation}
where $a$ and $b$ are complex numbers. The adjoint or regular matrix
representation of this operator is
\begin{equation}
\mathbf{H}=\left(
\begin{array}{llll}
0 & -ai & \left( \omega _{1}^{2}+\omega _{2}^{2}\right) i & 0 \\
0 & 0 & 0 & b\omega _{1}^{2}\omega _{2}^{2}i \\
-i & 0 & 0 & 0 \\
0 & 0 & ai & 0
\end{array}
\right) .  \label{eq:H_matrix}
\end{equation}
Its eigenvalues are the square roots of
\begin{equation}
\xi _{\pm }=\frac{\omega _{1}^{2}+\omega _{2}^{2}\pm \sqrt{4a^{2}b\omega
_{1}^{2}\omega _{2}^{2}+\left( \omega _{1}^{2}+\omega _{2}^{2}\right) ^{2}}}{%
2},  \label{eq:frequencies_square}
\end{equation}
and are real provided that
\begin{equation}
-\frac{\left( \omega _{1}^{2}+\omega _{2}^{2}\right) ^{2}}{4\omega
_{1}^{2}\omega _{2}^{2}}<a^{2}b<0.  \label{eq:a,b_conditions}
\end{equation}
The two possibilities $a^{2}>0$, $b<0$ or $a^{2}<0$, $b>0$ lead to the
Hamiltonians (\ref{eq:H_PU}) or (\ref{eq:H_PU_mod}), respectively. If we
choose $a^{2}b=-1$ and $\omega _{1}>\omega _{2}$ the resulting frequencies $%
\lambda _{1}^{2}=\omega _{1}^{2}$ and $\lambda _{2}^{2}=\omega _{2}^{2}$ are
related to the corresponding ladder operators
\begin{eqnarray}
Z_{1} &=&c_{1}\left[ \frac{\omega _{2}^{2}y}{a}+p_{x}-i\frac{\left( \omega
_{1}^{2}x+ap_{y}\right) }{\omega _{1}}\right] ,  \nonumber \\
Z_{2} &=&c_{2}\left[ \frac{\omega _{1}^{2}y}{a}+p_{x}-i\frac{\left( \omega
_{2}^{2}x+ap_{y}\right) }{\omega _{2}}\right] ,  \nonumber \\
Z_{3} &=&c_{3}\left[ \frac{\omega _{1}^{2}y}{a}+p_{x}+i\left( \omega _{2}x+%
\frac{ap_{y}}{\omega _{2}}\right) \right] ,  \nonumber \\
Z_{4} &=&c_{4}\left[ \frac{\omega _{2}^{2}y}{a}+p_{x}+i\left( \omega _{1}x+%
\frac{ap_{y}}{\omega _{1}}\right) \right] ,  \label{eq:Z_j_PU}
\end{eqnarray}
where $c_{i}$, $i=1,2,3,4$ are arbitrary real numbers. The only nonvanishing
commutators are
\begin{eqnarray}
\left[ Z_{1},Z_{4}\right] &=&\sigma _{1}=2c_{1}c_{4}\frac{\omega
_{1}^{2}-\omega _{2}^{2}}{\omega _{1}},  \nonumber \\
\lbrack Z_{2},Z_{3}] &=&\sigma _{2}=2c_{2}c_{3}\frac{\omega _{2}^{2}-\omega
_{1}^{2}}{\omega _{2}}.  \label{eq:[Zi,Zj]_PU}
\end{eqnarray}
Note that even these commutators vanish (breaking of the commutator
relation) in the case of equal frequencies which leads to an apparent
reduction of the number of degrees of freedom\cite{MD05,S06}. By means of
equation (\ref{eq:H(Z_j)}) we obtain
\begin{equation}
H=\frac{\omega _{1}}{\sigma _{1}}Z_{4}Z_{1}+\frac{\omega _{2}}{\sigma _{2}}%
Z_{3}Z_{2}+\frac{\omega _{2}-\omega _{1}}{2}=\frac{1}{2}\left( \frac{\omega
_{1}^{2}}{\omega _{1}^{2}-\omega _{2}^{2}}Z_{4}Z_{1}+\frac{\omega _{2}^{2}}{%
\omega _{2}^{2}-\omega _{1}^{2}}Z_{3}Z_{2}+\omega _{2}-\omega _{1}\right) .
\end{equation}
The different signs of the coefficients of $Z_{4}Z_{1}$ and $Z_{3}Z_{2}$
reveal the well known fact that the spectrum of this oscillator is unbounded
from above and below.

In the case of equal frequencies ($\omega _{1}=\omega _{2}=\omega $) the
matrix representation can be written in canonical Jordan form. Consider, for
example, the Hermitian version of the Pais-Uhlenbeck Hamiltonian (\ref
{eq:H_PU}) ($a=1$, $b=-1$) for which
\begin{equation}
\mathbf{H}=\left(
\begin{array}{cccc}
0 & -i & 2\omega ^{2}i & 0 \\
0 & 0 & 0 & -\omega ^{4}i \\
-i & 0 & 0 & 0 \\
0 & 0 & i & 0
\end{array}
\right) .  \label{eq:H_mat_ef}
\end{equation}
The matrix
\begin{equation}
\mathbf{P}=\left(
\begin{array}{cccc}
-\omega ^{2}i & 0 & \omega ^{2}i & 0 \\
\omega ^{3} & 3\omega ^{2} & \omega ^{3} & -3\omega ^{2} \\
\omega & 1 & \omega & -1 \\
-i & -\frac{2i}{\omega } & i & -\frac{2i}{\omega }
\end{array}
\right) ,  \label{eq:P_matrix}
\end{equation}
enables us to bring $\mathbf{H}$ into a canonical Jordan form by means of
the similarity transformation
\begin{equation}
\mathbf{P}^{-1}\mathbf{HP}=\left(
\begin{array}{cccc}
-\omega & 1 & 0 & 0 \\
0 & -\omega & 0 & 0 \\
0 & 0 & \omega & 1 \\
0 & 0 & 0 & \omega
\end{array}
\right) ,
\end{equation}
that exhibits two Jordan blocks of dimension $2$. In other words, the case
of equal frequencies $\omega _{1}=\omega _{2}=\omega $ corresponds to an
exceptional point where two pairs of eigenvectors of the matrix $\mathbf{H}$
coalesce and, consequently, only two eigenvectors remain linearly
independent. At this point the regular or adjoint matrix becomes defective
and can be transformed into a canonical Jordan form.

\section{Trivial quantization}

\label{sec:trivial}

The problem consists of obtaining a suitable quantum-mechanical Hamiltonian
that leads to the following fourth-order differential equation of motion for
the dynamical variables:
\begin{equation}
\frac{d^{4}}{dt^{4}}q+\left( \omega _{1}^{2}+\omega _{2}^{2}\right) \frac{%
d^{2}}{dt^{2}}q+\omega _{1}^{2}\omega _{2}^{2}q=0.  \label{eq:eq_mot}
\end{equation}
The solution is simple if we take into account that any quadratic function
of two coordinates and momenta already yields this differential equation as
shown in the preceding section and in an earlier communication\cite{F16b}.
Therefore, we only have to choose a suitable model from the large family of
such Hamiltonians. A well known textbook example will serve the purpose.
Consider two identical masses fixed to two opposite walls with two springs
with the same force constant $k$. Then we join the masses each other with a
third spring with a different force constant, say $k_{1}$. If the particles
move in only one dimension then the system will oscillate with two different
frequencies related to the two normal modes of vibration. Obviously, there
is no problem in quantizing this model and we will obtain a Hamiltonian
operator with a spectrum bounded from below. The case of equal frequencies
is obtained when $k_{1}=0$ so that the system becomes a pair of two
identical uncoupled harmonic oscillators. There is no problem in quantizing
this classical model either and we again obtain a spectrum that is bounded
from below. In both cases (different or equal frequencies) the
eigenfunctions will be square integrable.

For simplicity we consider a dimensionless equation. If we express the force
constants $k$ and $k_{1}$ in terms of the frequencies $\omega _{1}$ and $%
\omega _{2}$ of the two normal modes the Hamiltonian becomes
\begin{equation}
H=\frac{1}{2}\left( p_{1}^{2}+p_{2}^{2}\right) +\frac{\omega _{1}^{2}}{4}%
\left( x_{1}-x_{2}\right) ^{2}+\frac{\omega _{2}^{2}}{4}\left(
x_{1}+x_{2}\right) ^{2}.  \label{eq:H_diff}
\end{equation}
Note that
\begin{equation}
P(\lambda )=\lambda ^{4}-\left( \omega _{1}^{2}+\omega _{2}^{2}\right)
\lambda ^{2}+\omega _{1}^{2}\omega _{2}^{2},
\end{equation}
that is consistent with equation (\ref{eq:eq_mot}). This Hamiltonian is
separable
\begin{equation}
H=\frac{1}{2}\left( p_{x}^{2}+p_{y}^{2}\right) +\frac{\omega _{1}^{2}}{2}%
x^{2}+\frac{\omega _{2}^{2}}{2}y^{2},  \label{eq:H_diff_2}
\end{equation}
in terms of the coordinates
\begin{equation}
x=\frac{1}{\sqrt{2}}\left( x_{1}-x_{2}\right) ,\;y=\frac{1}{\sqrt{2}}\left(
x_{1}+x_{2}\right) .
\end{equation}
Therefore, the spectrum is bounded from below
\begin{equation}
E_{n_{1}n_{2}}=\omega _{1}\left( n_{1}+\frac{1}{2}\right) +\omega _{2}\left(
n_{2}+\frac{1}{2}\right) ,\;n_{1},n_{2}=0,1,\ldots ,  \label{eq:spect_diff}
\end{equation}
and the eigenfunctions $\psi _{n_{1}n_{2}}\left( x_{1},x_{2}\right) =\varphi
_{n_{1}}(\omega _{1},x_{1})\varphi _{n_{2}}(\omega _{2},x_{2})$, where $%
\varphi _{n}(\omega ,x)$ is a Harmonic-oscillator eigenfunction, are square
integrable.

The annihilation operators are
\begin{eqnarray}
a_{1} &=&\frac{\sqrt{\omega _{1}}}{2}\left( x_{1}-x_{2}\right) +\frac{i}{2%
\sqrt{\omega _{1}}}\left( p_{1}-p_{2}\right) ,  \nonumber \\
a_{2} &=&\frac{\sqrt{\omega _{2}}}{2}\left( x_{1}+x_{2}\right) -\frac{i}{2%
\sqrt{\omega _{2}}}\left( p_{1}+p_{2}\right) ,  \label{eq:annihi}
\end{eqnarray}
and the creation ones their adjoints. In terms of these operators the
Hamiltonian reads
\begin{equation}
H=\omega _{1}a_{1}^{\dagger }a_{1}+\omega _{2}a_{2}^{\dagger }a_{2}+\frac{1}{%
2}\left( \omega _{1}+\omega _{2}\right) .  \label{eq:H_diff_boson}
\end{equation}

Obviously, the case of equal frequencies does not offer any difficulty
because by construction the problem reduces to two uncoupled oscillators
\begin{equation}
H=\frac{1}{2}\left( p_{1}^{2}+p_{1}^{2}\right) +\frac{\omega ^{2}}{2}\left(
x_{1}^{2}+x_{2}^{2}\right) ,
\end{equation}
with spectrum
\begin{equation}
E_{n_{1}n_{2}}=\omega \left( n_{1}+n_{2}+1\right) .
\end{equation}

We call this solution to the problem of the quantization of the fourth-order
differential equation (\ref{eq:eq_mot}) trivial because it can be reduced to
two dynamical differential equations of second order that are known to offer
no difficulty. More precisely, the coordinates $x$ and $y$ satisfy
\begin{equation}
\ddot{x}+\omega _{1}^{2}x=0,\;\ddot{y}+\omega _{2}^{2}y=0.
\end{equation}

\section{Further comments and conclusions}

\label{sec:conclusions}

The algebraic method enables us to reduce the discussion of a quadratic
Hamiltonian to the analysis of its regular or adjoint matrix representation.
In this way we can easily elucidate several features of the spectrum of the
operator without solving its eigenvalue equation explicitly. It is, for
example, quite straightforward to determine whether the spectrum is real or
not. In this paper, in particular, we have stressed and exploited the fact
that any quadratic function of $K$ coordinates and their conjugate momenta
leads to a differential equation of order $2K$ for any of those dynamical
variables\cite{F16b}. As far as we are aware, this feature of the quadratic
Hamiltonians has not been taken into consideration in earlier studies of the
subject.

The regular or adjoint matrix representation of the operator also reveals
that at an exceptional point the matrix becomes defective (that is to say,
it has less than $2K$ linearly independent eigenvectors). At such point
there is a commutator breaking; that is to say, a pair of creation and
annihilation operators for the same degree of freedom commute. The reason is
that there is a one-to-one correspondence between the eigenvectors of the
adjoint or regular matrix representation and the creation and annihilation
(or ladder) operators. When two eigenvectors coalesce at an exceptional
point, the corresponding creation and annihilation operators are found to
commute.

At the exceptional point the defective matrix can be converted, by means of
a similarity transformation, into a canonical Jordan form (a matrix that
exhibits Jordan blocks). In the case of the Pais-Uhlenbeck oscillator this
phenomenon takes place at the equal-frequency limit. Although some of these
results are well known and have been discussed earlier, it has been our
purpose to show that the algebraic method enables us to study them in a
simple and unified way.

With respect to the problem posed by the quantization of the fourth-order
differential equation for the dynamical variables we have also shown that it
is possible to obtain a quantum-mechanical quadratic Hamiltonian with a
bounded-from-below spectrum and square-integrable eigenfunctions in the
cases of different as well as equal frequencies.

\end{document}